\documentclass[showpacs,aps,twocolumn]{revtex4}
\usepackage{graphicx}
\usepackage{amsfonts}
\usepackage{amsmath}
\usepackage{amssymb}
\usepackage{multirow}
\usepackage{braket}
\usepackage[usenames,dvipsnames]{color}

\begin{document}
\bibliographystyle{apsrev}

\title{Entropy of random symbolic high-order bilinear Markov chains}

\author{S.~S.~Melnik and O.~V.~Usatenko}

\affiliation{O. Ya. Usikov Institute for Radiophysics and
Electronics Ukrainian Academy of Science, 12 Proskura Street, 61805
Kharkiv, Ukraine}
\begin{abstract}
The main goal of this paper is to develop an estimate for the
entropy of random stationary ergodic symbolic sequences with
elements belonging to a finite alphabet. We present here the
detailed analytical study of the entropy for the high-order Markov
chain in the bilinear approximation. The appendix contains a short
comprehensive introduction into the subject of study.
\end{abstract}

\pacs{05.40.-a, 87.10+e, 07.05.Mh}

\maketitle

In the paper \cite{DNA16Mus}, we have presented results of our study
for the entropy of random long-range correlated symbolic sequences
with elements belonging to a finite alphabet. As a plausible model,
we have used the high-order additive Markov chain. Supposing that
the correlations between random elements of the chain are weak we
have expressed the conditional entropy of the sequence by means of
symbolic pair correlation functions. Here we present detailed
analytical calculations for the entropy of the high-order Markov
chain in the \emph{bilinear} approximation. The appendices contain a
comprehensive introduction in the matter of high-order Markov
chains.

\section{High-order Markov chains}

Consider a semi-infinite random stationary ergodic sequence
\begin{equation}\label{ranseq}
 \mathbb{A}= a_{0}, a_{1},a_{2},...
\end{equation}
 of symbols
(letters) $a_{i}$ taken from the finite alphabet
\begin{equation}\label{alph}
 \mathcal{A}=\{\alpha^1,\alpha^2,...,\alpha^m\},\,\, a_{i}\in \mathcal{A},\,\, i \in
\mathbb{N}_{+} = \{0,1,2...\}.
\end{equation}
We use the notation $a_i$ to indicate a position of the symbol $a$
in the chain and the notation $\alpha^k$ to stress the value of the
symbol $a\in \mathcal{A}$.

We suppose that the symbolic sequence $\mathbb{A}$ is a
\textit{high-order} \emph{Markov chain}~\cite{Raftery,
Seifert,Shields}. Such sequences are also referred to as multi- or
$N$-step~\cite{RewUAMM,MUYaG,UYa}, or \emph{categorical}~\cite{Hoss}
Markov's chains. One of the most important and interesting
application of the symbolic sequences is the probabilistic language
model specializing in predicting the next item in the sequence by
means of $N$ previous known items. There the Markov chain is known
as the $N$-\emph{gram model}.

The sequence $\mathbb{A}$ is the $N$-step Markov's chain if it
possesses the following property: the probability of symbol~$a_i$ to
have a certain value $\alpha^k \in \mathcal{A} $ under condition
that \emph{all} previous symbols are given depends only on $N$
previous symbols,
\begin{eqnarray}\label{def_mark}
&& P(a_i=\alpha|\ldots,a_{i-2},a_{i-1})\\[6pt]
&&=P(a_i=\alpha|a_{i-N},\ldots,a_{i-2},a_{i-1}).\nonumber
\end{eqnarray}
Sometimes the number $N$ is also referred to as the \emph{order} or
the \emph{memory length} of the Markov chain.

\section{Conditional entropy}

To estimate the conditional entropy of stationary sequence
$\mathbb{A}$ of symbols $a_{i}$ one could use the Shannon definition
\cite{Shan} for entropy per block of length $L$,
\begin{eqnarray} \label{entro_block}
H_{L}=-\sum_{a_{1},...,a_{L} \in \mathcal{A}} P(a_{1}^{L})\log_{2}
P(a_{1}^{L}).
\end{eqnarray}
Here $P(a_{1}^{L}) =P(a_{1},\ldots,a_{L})$ is the probability to
find $L$-word $a_{1}^{L}$ in the sequence; hereafter we use the
concise notation $a_{i-N}^{i-1}$ for $N$-word $a_{i-N},...,a_{i-1}$.
Instead the term \emph{word} one often uses the words: subsequence,
string or tuple. The conditional entropy, or the entropy per symbol,
is given by
\begin{eqnarray} \label{entro_diff}
h_{L}=H_{L+1} - H_{L}.
\end{eqnarray}
This quantity specifies the degree of uncertainty of $(L+1)$th
symbol occurring and measures the average information per symbol if
the correlations of $(L+1)$th symbol with preceding $L$ symbols are
taken into account. The conditional entropy $h_L$ can be represented
in terms of the conditional probability function
$P(a_{L+1}|a_{1}^{L})$, 
\begin{eqnarray} \label{Entro_Bin}
h_L=\!\!\sum_{a_{1},...,a_{L} \in \mathcal{A}}\!\! P(a_{1}^{L})
h(a_{1}^{L}) = \overline{ h(a_{1}^{L})} ,
\end{eqnarray}
where $h(a_{1}^{L})$ is the amount of information contained
in the $(L+1)$th symbol of the sequence conditioned on $L$ previous
symbols $a_1^L$,
\begin{eqnarray}
   h(a_{1}^{L}) = - \!\!\sum_{a_{L+1} \in \mathcal{A}}\!\!
P(a_{L+1}|a_{1}^{L})\log_2 P(a_{L+1}|a_{1}^{L}).
    \label{siL}
\end{eqnarray}
The source entropy (or Shannon entropy) is the conditional entropy
at the asymptotic limit, $h=\lim_{L \rightarrow \infty}h_{L}$. This
quantity  measures the average information per symbol if {\it all}
correlations, in the statistical sense, are taken into account, cf.
with \cite{Grass}, Eq.~(3).

Supposed stationarity of the random sequence under study together with decay of correlations,
\emph{$C_{\alpha,\beta}(r\to\infty) \rightarrow 0$}, see below
definition~(\ref{def_cor1}), lead, according to the Slutsky
sufficient conditions~\cite{Slu}, to the mean-ergodicity.
Due to the ergodicity, the ensemble
average of any function $f(a_{r_1},a_{r_1+r_2},\ldots ,
a_{r_1+\ldots+r_{s}})$ of $s$ arguments defined on the set $A$ of
symbols can be replaced by the statistical (arithmetic, Ces\`{a}ro's) average over
the chain.  This latter
property is very useful in numerical calculations since the
averaging can be done over the sequence and the ensemble averaging
can be avoided. Therefore, in our numerical as well as analytical
calculations, we always apply averaging over the length of the
sequence as it is implied in Eq.~(\ref{Entro_Bin}).

\section{Correlation functions and words}

If the sequence, statistical properties of which we would like to
analyze, is given, the \emph{conditional probability distribution
function} (CPDF) of $N$th order can be found by a standard method
(written below for subscript $i=N+1$)
\begin{equation}\label{soglas}
P(a_{N+1}=\alpha^k|a_{1}^{N})=\frac{ P(a_{1}^{N},\alpha^k) } {
P(a_{1}^{N})},
\end{equation}
where $P(a_{1}^{N},\alpha^k)$ and $P(a_{1}^{N})$ are the
probabilities  of the $(N+1)$-subsequence $a_{1}^{N},\alpha^k$ and
$N$-subsequence $a_{1}^{N}$ occurring, respectively.

The conditional probability function completely determines \emph{all
statistical properties} of the random chain and the method of its
generation. Equation~(\ref{soglas}) says that the CPDF is determined
if we know the probability of $(N+1)$-words occurring -- the words
containing $(N+1)$ symbols without omissions among their indexes.
Obviously, the average number of some word $a_1^L$ occurring in
whole sequence exponentially decreases with the word length $L$. Let us
evaluate the length $L_{max}$ of word, that occurs on average one
time. For given length $M$ of weakly correlated sequence with fixed
dimension $m$ of the alphabet this length, evidently, is equal to
$L_{max}\thickapprox \ln M/\ln m $.

To make this evaluation more precise we should take into account
that the correlations decrease the number of \emph{typical} words
that one can encounter in the sequence and this phenomenon increases
the length $L_{max}$. From the famous result of the theory of
information, known under the name of the Shennon-McMillan-Breiman
theorem~\cite{Cover}, it follows
\begin{equation}\label{Lmax}
L_{max}\sim \frac{\log_2 M}{h},
\end{equation}
where $h$ is the conditional entropy  per letter of sequence under
condition that all correlations are taken into account. This is a
crucial point, because the correlation lengths of natural sequences
of interest are usually of the same order as the sequence length,
whereas the last inequality can only be fulfilled for the maximal
lengths of the words $L_{max} \lesssim 10$.

The words of the length $L\ll L_{max}$ are well represented in the
sequence, so that one can use the statistical approach to these
objects and calculate directly the probabilities of their occurrence in the
chain. By contrast, the statistics of longer words, $L \gtrsim
L_{max}$, are insufficient and the whole sequence for such words is
not anymore probabilistic object. Some papers devoted to this
question even put under doubt the correctness of the
notion of ``finite random sequence''~\cite{Doyle,Uspen}.

So, if the correlation length $R_c$ of sequence is less than $L_{max}$,
then the random sequence should be considered as
quasi-ergodic because the words of the length $L \leq R_c < L_{max}$
provide statistically meaningful information for reconstructing the
conditional probability function of the sequence.

We meet a completely different situation when $ L_{max} < R_c$. In
this case the statistical properties of the studied sequence can be
reconstructed only up to the length of order $ L \ll  L_{max}$.
Statistically important information on the properties of the
sequence in the interval $L_{max} <L<R_c$ is inaccessible in the
frame of discussed likelihood estimation method.

For simplicity of further qualitative consideration let us fix our
attention on the pair correlation function only. If we know statistics of $(N+1)$-words, we also know the
correlation function for $r\leqslant N$. Nevertheless, for the given sequence of length
$M$, we can calculate $C_{\alpha\beta}(r)$ at $r$, which is of order
of $M$. Really, for a weakly correlated sequence, the probability
$P(a_i=\alpha, a_{i+r}=\beta)$ to have the pair of letters $\alpha$
and $\beta$ at the distance $r$ is equal to $p_\alpha p_\beta$. This
quantity determines the number of pairs in the hole sequence. The
number of pairs is a slowly decreasing function of $r$. As above we
can evaluate the distance $r_{max}$ between the pair of letters that
occurs on average one time. Thus, from definition of the probability
$P(a_i=\alpha, a_{i+r}=\beta)=N_{ab}/(M-r)$, where $N_{ab}$ is the
number of pairs $ab$ in the interval $M-r$, we have
$r_{max}\thicksim M - 1/(p_\alpha p_\beta)$. It is clear that
$r_{max}$ can be much greater than $L_{max}$.

For $k$-order correlation functions or $k$-words, the estimation is
$r^{(k)}_{max}\thicksim M - 1/(p_{\alpha_1}... p_{\alpha_k})$.

Let us note that in the frames of both methods we cannot take into
account the correlation functions of order exceeding $L_{max}$. This
quantity determines both the maximal length of words, without or
with omission of symbols among them in the sequence (in mathematics
such sets are known under the name of \emph{cylinder} ones), and the
maximal order of correlation functions, which can be used to
describe statistical properties of the sequence.
In the general case the differences among the arguments of the correlation functions are
limited by $r^{(k)}_{max}$. The information about the region
$L_{max} \lesssim L \ll \texttt{min}(R_c,r_{max})$ is introduced in
consideration by means of the memory functions, which are expressed
through the correlation functions.

A method that allows us to use the information on the symbols spaced
by a distance $ r \ll \texttt{min}(R_c,r_{max})$, not only in
narrower region with $r \ll L_{max}$, is connected with the
high-order additive and bilinear Markov chains, a construction
proposed in Ref.~\cite{MUYaG,muya,arxiv}.

\begin{widetext}
\section{Memory and correlation functions}

The \emph{bilinear} Markov chain is determined by the conditional
probability distribution function of the form

\begin{eqnarray}\label{CPF}
  P(a_i=\alpha|a_{i-N}^{i-1})& =& p_\alpha + \sum_{r=1}^N \sum_{\beta \in
  \mathcal{A}}
  F_{\alpha\beta}(r)[\delta({a_{i-r}, \beta}) - p_\beta] \\ \nonumber
  &+& \sum_{1=r_1<r_2}^N \sum_{\beta, \gamma \in \mathcal{A}} F_{\alpha \beta \gamma}
  (r_1, r_2)\left[[\delta({a_{i-r_1}, \beta}) - p_\beta][\delta({a_{i-r_2},
  \gamma}) - p_\gamma] - C_{\gamma\beta}(r_2 - r_1)\right],
\end{eqnarray}
\end{widetext}
where $p_{\alpha}$ is the relative number of symbols $\alpha$ in the
chain, or their probabilities of occurring,
\begin{equation}\label{a-av}
p_{\alpha}=\overline{\delta(a_i , \alpha)}.
\end{equation}
Here $\delta(.,.)$ is the Kronecker delta-symbol. The  quantities
$F_{\alpha\beta}(r)$ and $F_{\alpha \beta \gamma}
  (r_1, r_2)$ are the so called \emph{memory functions}. In Appendix A some
  suggestions on the
form of Eq.~(\ref{CPF}) and its properties are presented.

As a rule, the statistical properties of random sequences are
determined by correlation functions. The \emph{symbolic correlation functions}
of the $k$th order are
given by the following expression,
\begin{eqnarray}\label{def_cor1}
&&C_{\beta^1,...\beta^k}(r_1,r_2,\ldots,r_{k-1}) \\ [6pt]
=&&\overline{ [\delta(a_{0},\beta^1)-p_{\beta^1}]\ldots
[\delta(a_{r_1+\ldots+r_{k-1}},\beta^k)-p_{\beta^k}]}.\nonumber
\end{eqnarray}
The $\overline{overline}$ means a statistical average over an
ensemble of sequences.
Note that in some sense symbolic
correlation functions-matrices are more general construction than
numeric correlation functions. They can describe in more detail even
numeric sequences.

There were suggested two methods for finding the memory functions of
a sequence with a known correlation functions. The first
one~\cite{muya} is based on the minimization of the ``distance''
between the conditional probability function, containing the
sought-after memory function, and the given sequence $\mathbb{A}$ of
symbols with a known correlation function,
\begin{equation}\label{Dev_eq40}
Dist = \overline{\sum_\alpha[\delta(a_{i},\alpha) - P(a_i=\alpha|a_{i-N}^{i-1})]^2}.
\end{equation}
In this equation a given sequence $\mathbb{A}$ is presented by the
Kronecker delta-function, the unknown parameters are the memory
functions of the CPDF.

The second method for deriving the equations connecting the memory
and correlation functions is a completely probabilistic
straightforward calculation analogous to that used
in~\cite{MUYaG,arxiv}. These equations, despite its simplicity, can
be analytically solved only in some particular cases: for one- or
two-step chains, the Markov chain with a step-wise memory function
and so on. To avoid the various difficulties in its solving we
suppose that correlations in the sequence are weak (in amplitude,
but not in length). An approximate solution for the memory function
allows one to obtain the following simple formulas
\begin{eqnarray}\label{Dev_eq41}
   F_{\alpha\beta}(r)\!=\!\frac{C_{\beta\alpha}(r)}
  {p_\beta}, \, F_{\alpha \beta \gamma}
  (r_1, r_2)\!=\!\frac{C_{\beta\gamma\alpha}
  (r_2 - r_1, r_1)}{p_\beta p_\gamma}.
\end{eqnarray}

Equation \eqref{CPF} together with Eq.~\eqref{Dev_eq41} provide a
tool for constructing weak correlated sequences with given pair and
third-order correlation functions~\cite{muya,arxiv}. Note that
$i$-independence of the function $P(a_i=\alpha|a_{i-N}^{i-1})$
provides homogeneity and stationarity of the sequences under
consideration; and finiteness of $N$ together with the strict
inequalities
\begin{equation}\label{ergo_m}
 0 < \!P(a_{i+N}\!=\alpha|a_{i}^{i+N-1})\! < 1, \, i \in \mathbb{N}_{+} = \{0,1,2...\}
\end{equation}
provides, according to the Markov theorem (see, e.g.,
Ref.~\cite{shir}), ergodicity of the sequences.

\section{Entropy of Bilinear chain}

The conditional probability distribution function
$P(a_i=\alpha|a_{i-N}^{i-1})$ determined by Eq.~\eqref{CPF} gives
the probability to have a symbol $a_i=\alpha$ after $N$-word
$a_{i-N}^{i-1}$. Nevertheless, we would like to know the conditional
entropy not only after $N$-word, but in all range of $L$-words. The
conditional probability $P(a_i=\alpha|a_{i-L}^{i-1})$ for a word of
length $L < N$ can be obtained in the first approximation in the
weak correlation parameter $\Delta_{\alpha}(L)$ from general
definition Eqs.~\eqref{def_mark} by means of a routine probabilistic
reasoning presented in Appendix B. The rule to obtain the CPDF
consists in replacement in Eq.~\eqref{CPF} $N\rightarrow L$
\begin{equation}\label{Entro_Markov2}
\!P(a_i=\alpha|a_{i-L}^{i-1}) \! = \! \left\{\begin{array}{l}
  P(a_i=\alpha|a_{i-N}^{i-1}),\qquad \,L \geqslant N \\[8pt]
 P(a_i=\alpha|a_{i-N}^{i-1})_{|_{N \rightarrow L}}, \, L < N.
\end{array}
\right.
\end{equation}
The first line follows from the markovian property of the CPDF, Eq.~
\eqref{def_mark}.

We suppose that the random sequence is weakly correlated and present
CPDF, Eq.~\eqref{CPF}, as
\begin{equation}\label{Delta0}
    P(a_i=\alpha|a_{i-N}^{i-1})=p_\alpha +
\Delta_\alpha(a_{i-N}^{i-1}),
\end{equation}
admitting that the strong inequalities
\begin{equation}\label{Delta}
  |\Delta_\alpha| \ll p_\alpha,
\end{equation}
are fulfilled.

Expanding the conditional entropy $h(a_1^L)$ in series with respect
to the small $\Delta$ we  have
\begin{equation}\label{h_Raw2}
    h \approx h_0 + \sum_{\alpha}\frac{\partial h}{\partial
    P(\alpha)}\Delta_\alpha + \frac{1}{2} \sum_{\alpha \beta}
    \frac{\partial^2 h}{\partial P(\alpha) \partial P(\beta)}\Delta_\alpha \Delta_\beta,
\end{equation}
where
\begin{eqnarray}\label{h_delta1}
  & h &= h(a_1^L), \,h_0 = -\sum_{\alpha \in \mathcal{A}} p_\alpha \log_2 p_\alpha,\,\\ \nonumber
 & P(\alpha)&= P(a_{L+1}=\alpha|a_1^L).
\end{eqnarray}

It is important to note that the number of independent variables of
the function $h$ is equal to $m-1$, because the sum,
\[
    \sum_{\alpha \in \mathcal{A}} P(a_{L+1}=\alpha|a_1^L) = 1,
\]
expressing the normalization condition of the CPDF, is fixed. By
this reason, it is convenient to present Eq.~\eqref{h_Raw2} in the
form,
\begin{widetext}
\begin{equation}\label{h_cond}
    h = -\frac{1}{\ln 2} \sum_{i=1}^{m-1} P(\alpha^i) \ln P(\alpha^i)
     -\frac{1}{\ln 2} (1 - \sum_{i=1}^{m-1} P(\alpha^i)) \ln (1 -
     \sum_{i=1}^{m-1} P(\alpha^i)),
\end{equation}
\end{widetext}
containing independent variables only. After
differentiation of Eq.~\eqref{h_cond} and substitution of obtained
derivatives (taken at the point $P(a_i=\alpha|a_{i-L}^{i-1}) =
p_{\alpha}$) in Eq.~\eqref{h_Raw2}, we have
\begin{equation}\label{h_cond2}
    h = h_0 -\frac{1}{\ln 2} \sum_{\alpha \in \mathcal{A}} \Delta_\alpha \ln p_\alpha -
    \frac{1}{2 \ln 2} \sum_{\alpha \in \mathcal{A}}
    \frac{\Delta^2_\alpha}{p_\alpha}.
\end{equation}
The quantities $\Delta_\alpha$ depend on a concrete $L$-word $a_1^L$
preceding the generated symbol $a_{L+1}=\alpha$. To obtain the
conditional entropy $h_L$ we should substitute  \eqref{h_cond2} into
 \eqref{Entro_Bin} to average it. As a result we have
\[
    h_L = \overline{h} = h_0 -\frac{1}{\ln 2} \sum_{\alpha \in \mathcal{A}}
    \overline{\Delta_\alpha} \ln p_\alpha -\frac{1}{2 \ln 2}
    \sum_{\alpha \in \mathcal{A}} \frac{\overline{\Delta^2_\alpha}}{p_\alpha}.
\]
Here the second term is equal zero, because $\Delta_\alpha$ is taken
at the ``equilibrium point'' $P(a_i=\alpha|a_{i-L}^{i-1}) =
p_{\alpha}$. As a result, we have
\begin{equation}\label{h_delta}
    h_L = h_0 -\frac{1}{2 \ln 2} \sum_{\alpha \in \mathcal{A}} \frac{\overline
    {\Delta^2_\alpha}}{p_\alpha}.
\end{equation}
\begin{widetext}
In the case of weak correlations, we can calculate the dispersion of
$\Delta_\alpha$ after expressing the memory functions by meas of
correlation functions,
\begin{eqnarray}\label{h_delta2}
  \Delta_\alpha &=& \sum_{r=1}^L \sum_\beta \frac{C_{\beta\alpha}(r)}
  {p_\beta}[\delta({a_{i-r}, \beta}) - p_\beta] \\ \nonumber
  &+& \sum_{r_1<r_2}^L \sum_{\beta, \gamma} \frac{C_{\beta,\gamma,\alpha}
  (r_2 - r_1, r_1)}{p_\beta p_\gamma} \{[\delta({a_{i-r_1}, \beta}) -
  p_\beta][\delta({a_{i-r_2}, \gamma}) - p_\gamma] - C_{\gamma\beta}(r_2 - r_1)\}.
\end{eqnarray}
Under calculation of $\overline{\Delta^2_\alpha}$ there appear three
terms. The first one is
\begin{eqnarray}\label{h_delta3}
  D_\alpha^{(11)} &=& \sum_{r,r'=1}^L \sum_{\beta, \beta'}
   \frac{C_{\beta\alpha}(r)}{p_\beta} \frac{C_{\beta'\alpha}(r')}
   {p_{\beta'}} \overline{[\delta({a_{i-r}, \beta}) - p_\beta][
      \delta({a_{i-r'}, \beta'}) - p_{\beta'}]} \\ \nonumber
  &=& \sum_{r,r'=1}^L \sum_{\beta, \beta'} \frac{C_{\beta\alpha}
  (r)}{p_\beta} \frac{C_{\beta'\alpha}(r')}{p_{\beta'}} C_{\beta \beta'}(r-r').
\end{eqnarray}
In the weak-correlation approximation, the main contribution in
$D_\alpha^{(11)}$ gives the term with the function $C_{\beta
\beta'}(r-r')$ taken at $r'=r$,
\begin{equation}\label{D11}
    D_\alpha^{(11)} = \sum_{r=1}^L \sum_\beta \frac{C_{\beta
    \alpha}^2(r)}{p_\beta},
\end{equation}
where we have taken into account $C_{\alpha
\beta}(0)=p_{\alpha}\delta(\alpha,\beta)-p_{\alpha} p_{\beta}, C_{
\beta,\beta}(0)\simeq p_{\beta} $.
In the same way, we obtain a contribution to $D_\alpha^{(12)}$:
\begin{equation}\label{D111}
D_\alpha^{(12)} = \sum_{r=1}^L \sum_{1=r_1<r_2}^L \sum_{\beta \beta'
\gamma}
    \frac{C_{\beta \alpha}(r) C_{\beta' \gamma \alpha}(r_1, r_2) C_{\beta'
    \gamma \beta}(r_2 - r, r_1 - r)}{p_\beta p_{\beta'} p_\gamma}.
\end{equation}
Since none of the correlators has zero arguments (in the third term
one of arguments may take a zero value, but others are not equal
zero) the hole expression is small with respect to the term
$D_\alpha^{(11)}$ in the limiting case of small correlarions.

For the last contribution $D_\alpha^{(22)}$ one gets:

\begin{eqnarray}\label{h_Delta}
  D_\alpha^{(22)} &=& \sum_{1=r_1<r_2}^L \sum_{1=r'_1<r'_2}^L
    \sum_{\beta \gamma \beta' \gamma'} \frac{1}{p_\beta p_\gamma p_{\beta'}
     p_{\gamma'}}C_{\beta \gamma \alpha}(r_2, r_1) C_{\beta' \gamma' \alpha}
     (r_2', r_1') \\ \nonumber
     &\times& \left[ C_{\beta \gamma \beta' \gamma'}(r_2, r_1, r_2', r_1')
     - C_{\gamma\beta}(r_2 - r_1)C_{\gamma'\beta'}(r_2' - r_1')
     \right].
\end{eqnarray}
\end{widetext} where all arguments of four-order correlation
functions, for convenience of further analysis, are expressed by
means of the distances between the current and generated $a_i$
symbols.

The four-order correlation function takes a maximal value under
condition of coincidence of two pair of its arguments: $r_1 = r_1'$,
$r_2 = r_2'$, when it takes the value $C_{\beta \beta'}(0) C_{\gamma
\gamma'}(0)$. The second term in the square bracket is small with
respect to the first one. From here one obtains for
$D_\alpha^{(22)}$:
\[
    D_\alpha^{(22)} = \sum_{1=r_1 < r_2}^L \sum_{\beta \gamma}
    \frac{C^2_{\beta \gamma \alpha}(r_2, r_1)}{p_\beta p_\gamma}.
\]

So, taking into account the obtained above expressions  for
$D_\alpha^{(11)}$ è $D_\alpha^{(22)}$,  the  average deviation
$\overline{\Delta^2_\alpha}$ is presented by meas of second and
third-order correlation functions. Substituting their expressions
in Eq.~\eqref{h_delta}, we get the
desired result for the conditional entropy in the limiting case of
weak second and third order correlations,
\begin{equation}\label{h}
    h_L = h_0 - \frac{1}{2 \ln 2} \left[ \sum_{r=1}^L \sum_{\alpha \beta}
    \frac{C_{\beta \alpha}^2(r)}{p_\alpha p_\beta} \!+\!\!\! \sum_{r_1 < r_2}^L
    \!\!\sum_{\alpha \beta \gamma} \!\!\frac{C^2_{\beta \gamma \alpha}(r_2, r_1)}
    {p_\alpha p_\beta p_\gamma} \right].
\end{equation}
If the length of block exceeds the memory length, $L>N$, the
conditional probability $P(a_{i}=\alpha|a_{i-L}^{i-1})$ depends only
on $N$ previous symbols, see Eqs.~(\ref{def_mark})
and~(\ref{Entro_Markov2}). Then, it is easy to show
from~\eqref{Entro_Bin} that the conditional entropy remains constant
at $ L \geqslant N$. This property can be used for the numeric
definition of the sequence memory length.

\appendix

\section{}

Accepting definition~\eqref{def_mark} of the high-order Markov chain
as a starting point, we present in this section different models for
the conditional probability distribution function (CPDF) of symbolic
random sequences. It is helpful to present it as a \emph{finite}
polynomial series containing $N$ Kronecker delta-symbols,
\begin{eqnarray} \label{Gen-P}
P(a_i=\alpha|a_{i-1},\ldots,a_{i-N}) \\ \nonumber
=\sum_{\beta_1 \ldots \beta_N \in \mathcal{A}} F_{\alpha;\beta_1 \ldots \beta_N}
\prod_{r=1}^N \delta(a_{i-r},\beta_r).
\end{eqnarray}
This form of CPDF express some ``independence'' of the random
variables $a$ and the spatial coordinates $i$. The function
$F_{\alpha;\beta_1 \ldots \beta_N}$ is referred to as the
\textit{generalized memory function} and the Kronecker delta-symbols
play the role of the \emph{indicator} function of random variable
$a_{i-r}$ converting symbols to numbers $0$ or $1$.

Let us decouple the memory function
$F_{\alpha;\beta_1 \ldots \beta_N}$ and present it in the form
of the sum of \emph{memory functions} of $k$th \emph{order},
$F^{(k)} = F_{\alpha;\beta_1 \ldots \beta_k}(r_1, \ldots, r_k)$,
\begin{equation} \label{DecoupF}
F_{\alpha;\beta_1 \ldots \beta_N} = \sum_{k=0}^N \sum_{\lbrace r_1,\ldots r_k \rbrace}
F_{\alpha;\beta_1 \ldots \beta_k}(r_1, \ldots, r_k),
\end{equation}
where all symbols $r_s$ at the right hand side of
Eq.~\eqref{DecoupF} are different, ordered,
\begin{equation} \label{DecoupF2}
 1 \leqslant r_1 < r_2 < \ldots < r_k \leqslant N,
\end{equation}
and contain all different subsets $\{r_1,\! \ldots, r_k\}$ picked
out from the set $\{1, \ldots, N\}$. The coordinates $r_s$ of the
memory function $F_{\alpha;\beta_1 \ldots \beta_k}(r_1, \ldots, r_k)$ indicate positions of elements
$a_{i-r_s}$ taking on the values $\beta_s$.

\textbf{Uncorrelated sequence } known also as a \emph{discrete white
noise} or the \emph{Bernoulli scheme} is defined by the
past-independent function
\begin{equation}\label{Dev_eq2'}
P(a_i=\alpha|a_{i-N}^{i-1})=P(a_i=\alpha) = p_{\alpha}.
\end{equation}
It is the simplest and most well studied random sequence. This
sequence can be obtained by taking into account $r$-independent
function of zero order $F_{\alpha}$.

\textbf{One-step Markov chain} or the \emph{ordinary markovian
chain} is given by the two-parameter transition probability matrix
function $p_{\alpha\beta}$,
\begin{equation}\label{Dev_eq2}
P(a_i=\alpha|a_{i-N}^{i-1})=P(a_i=\alpha|a_{i-1}=\beta)=p_{\alpha\beta}.
\end{equation}
The CPDF of this sequence is obtained from Eq.~\eqref{DecoupF} by
taking into account two terms $F^{(0)}=F_{\alpha}$ and
$F^{(1)}=F_{\alpha;\beta}(1)$.

\textbf{Additive high-order Markov chain.}  For this random sequence
the CPDF takes on the ``linear form'' with respect to the Kronecker
delta-symbols,
\begin{eqnarray} \label{Dev_eq3}
&&P_{add}(a_i=\alpha|a_{i-N},\ldots,a_{i-2},a_{i-1})  \nonumber \\
\!&&\!= p_{\alpha}+\sum_{r=1}^N \! \sum_{\beta \in \mathcal{A}}\!\!
F_{\alpha \beta}(r)[\delta(a_{i-r},\beta)-p_{\beta}].
\end{eqnarray}
The additivity means that the previous symbols $a_{i-N}^{i-1}$ exert
an independent effect on the probability of the symbol $a_i=\alpha$
occurring. The conditional probability function in
form~(\ref{Dev_eq3}) can reproduce correctly the pair (two-point) correlations in the chain. The higher-order correlators
and all correlation properties of higher orders cannot be reproduced correctly by means of
the memory function $F_{\alpha \beta}(r)$. To understand better how
we can obtain Eq.~(\ref{Dev_eq3}), let us consider its simpler
forms.

\textbf{Binary additive high-order Markov chain with step-wise
memory function}. For the binary state space $a_i\in \{0, 1\}$ the conditional probability distribution function
to have the symbol ``1'' after $N$-word containing $k$ unities, is supposed
to be of the form,
\[
P(a_{N+1}=1\mid \underbrace{11\dots 1}_{k}\;\underbrace{00\dots 0}
_{N-k})
\]
\begin{equation}
=\frac{1}{2}+\mu (\frac{2k}{N}-1). \label{14}
\end{equation}
Here the correlation parameter $\mu $ belongs to the region
determined by inequality $-1/2< \mu <1/2$. It is exactly solvable
model \cite{UYa,UYaKM-03,MUYaAM-06-AllMemStepCor}.

\textbf{Binary additive high-order Markov chain.} It is the more
complicated model. Its CPDF  of random variables $a_i\in \{0, 1\}$,
the probability of symbol~$a_i$ to have a value $1$ under the
condition that $N$ previous symbols $a_{i-N}^{i-1}$ are given, is of
the following form~\cite{muya,RewUAMM},
\begin{equation} \label{prob1}
 P(a_i=1|a_{i-N}^{i-1})) = \bar{a} + \sum_{r=1}^N F(r)(a_{i-r} -
\bar{a}).
\end{equation}
Here $\bar{a}$ is the relative average number of unities in the
sequence. The representation of Eq.~(\ref{prob1}) in this form is
followed from the simple identical equalities, $a^2=a$ and $f(a) = a
f(1) + (1-a) f(0)$,  for an arbitrary function $f(a)$ determined on
the set $a\in \{0, 1\}$. If $F(r)$ is constant and $\bar{a}=1/2$ we return to the
previous model of the binary additive high-order Markov chain with
step-wise memory function.

For $P(0|.)$ we have from Eq.~\eqref{prob1},
\begin{eqnarray} \label{prob0}
P(a_i&=&0|a_{i-N}^{i-1}) =1-
P(1|a_{i-N}^{i-1})\nonumber \\
&=&1-\bar{a}-\sum_{r=1}^N F(r)(a_{i-r} - \bar{a}).
\end{eqnarray}
This two expressions are not symmetric with respect to the change $0
\leftrightarrows 1$ of generated symbol $a_i$. Let us show that
Eqs.~(\ref{prob1}) and~(\ref{prob0}) can be presented in the
symmetric form,
\begin{equation}\label{prob2}
P(a_i=\alpha|a_{i-N}^{i-1})\!=\! p_{\alpha}+\!\!\sum_{r=1}^N \!
\sum_{\beta \in \{0,1\}}\!\!\! F_{\alpha
\beta}(r)[\delta(a_{i-r},\beta)-p_{\beta}].
\end{equation}

Taking into account the definitions $p_1=\bar{a}$, $p_0=1-\bar{a}$,
using the  evident equalities $\delta(a_{i-r},0)=1-a_{i-r}$,
$\delta(a_{i-r},1)=a_{i-r}$ and putting $ F_{1 1}(r)- F_{1 0}(r)
=F_{00}(r)- F_{01}(r)= F(r)$  we easily obtain Eqs.~(\ref{prob1})
and (\ref{prob0}).

We should replace $\alpha,\beta \in \{0,1\}$ in Eq.~(\ref{prob2}) by
$\alpha,\beta \in \mathcal{A}$ to obtain Eq.~(\ref{Dev_eq3}).

Note, there is no one-to-one correspondence between the memory
function $F_{\alpha \beta}(r)$ and the conditional probability
function $P(a_i=\alpha|a_{i-N}^{i-1})$. Indeed, it is easy to see
that, in view of Eqs.~(\ref{Dev_eq2}) and~(\ref{a-av}), the
renormalized memory function $F'_{\alpha \beta}(r)=F_{\alpha
\beta}(r)+\varphi_{\alpha} (r)$ provides the same conditional
probability as $F_{\alpha \beta}(r)$.

\textbf{Bilinear high-order Markov chain.} The CPDF of this chain
Eq.~\eqref{CPF} is the direct generalization of Eq.~\eqref{Dev_eq3}.
More detailed explanation is given in Ref.~\cite{arxiv}.

\section{}

Here we prove Eq.~(\ref{Entro_Markov2}) using Eq.~(\ref{CPF})
as a starting point. From the
definition of the CPDF it follows
\begin{equation}\label{App1}
P(a_i=\alpha|W)=\frac{P(W,\alpha)} { P(W)}, \quad W =
a_{i-N+1}^{i-1}.
\end{equation}
Adding symbol $a_{i-N}=\beta$ to the word $(W,\alpha)$, we have
\begin{equation}\label{App2}
P(a_i=\alpha|W)=\frac{\sum_{\beta \in A}P(\beta,W,\alpha)} { P(W)}.
\end{equation}

Replacing here the probabilities $P(\beta,W,\alpha)$ by the CPDF
$P(a_i=\alpha|\beta,W)$ from the equation similar to that of
Eq.~(\ref{App1}),
\begin{equation}\label{App2'}
P(a_i=\alpha|\beta,W)=\frac{P(\beta,W,\alpha)} { P(\beta,W)},
\end{equation}
we obtain
\begin{equation}\label{App3}
P(a_i=\alpha|W)=\frac{1}{ P(W)}\sum_{\beta \in
A}P(\beta,W)\{P(a_i=\alpha|\beta,W)\} .
\end{equation}
This equation gives the CPDF $P(a_i=\alpha|W)$, based on the word
$W$ of the length $N-1$, my means of $N$-word. To prove the second
line of Eq.~(\ref{Entro_Markov2}), we should estimate two residual
terms. The first one,
\begin{eqnarray}
\frac{1}{P(W)}\sum_{\gamma\in A}F_{\alpha\gamma}(N)\sum_{\beta\in A}
 P(\beta,W)\left[ \delta(\beta,\gamma)-p_\gamma\right],
\label{App32}
\end{eqnarray}
is taken from additive part of the CPDF, and the second one does from its
bilinear part,
\begin{widetext}
\begin{equation}\label{CPF3}
\frac{1}{ P(W)}\sum_{\beta \in A}P(\beta,W)\sum_{r_1=1}^{N-1}
\sum_{\gamma, \rho} F_{\alpha \gamma \rho}(r_1, N)
\{[\delta({a_{i-r_1},\gamma}) - p_\gamma][\delta({\beta, \rho}) -
p_\rho] - C_{\rho\gamma}(N - r_1)\}.
\end{equation}
\end{widetext}

We obtain them after separation $ P(a_i=\alpha|a_{i-N+1}^{i-1})$
from the term $P(a_i=\alpha|a_{i-N}^{i-1})$. In the paper
\cite{DNA16Mus} it was shown that the term Eq.~\eqref{App32} is
small with respect to the others terms of the sum over $r\in
(1,...,N-1)$ in Eq.~\eqref{CPF}. So, we should show that the
term~\eqref{CPF3} also is small.

In Eq.~\eqref{CPF3}, let us consider the factor $\sum_{\beta \in
A}P(\beta,W) [\delta({\beta, \rho}) - p_\rho]$ and present it in the
form
\begin{equation}\label{App3'}
 P(\rho,W)(1-p_\rho)-P(\overline{\rho},W)p_\rho,
 \end{equation}
where the symbol $\overline{\rho}$ stands for an event NOT-$\rho$.
It is intuitively clear that in the zero approximation in $\Delta$
(i.e., for uncorrelated sequence, when $P(\rho,W)\approx
P(\rho)P(W)$) this term equals zero,
$P(\rho,W)(1-p_\rho)-P(\overline{\rho},W)p_\rho\simeq
  P(W)[p_\rho (1-p_\rho)-p_{\overline{\rho}}p_{\rho}]$. In the next approximation
this term is of order of $\Delta$. These two statements can be
verified by using the condition of compatibility for the
Chapman-Kolmogorov equation (see, for example, Ref.~\cite{gar}),
\begin{equation}\label{App4}
P(a_{i-N+1}^{i})=\sum_{a_{i-N}\in A}P(a_{i-N}^{i-1})
P_{N}(a_i|a_{i-N}^{i-1}).
\end{equation}

The term in Eq.~\eqref{CPF3} containing $C_{\rho\gamma}(N - r_1)$ is
of the same order as considered above Eq.~\eqref{App3'}, because of
the inequality $N \neq r_1$ in Eq.~\eqref{CPF3} is fulfilled.

Hence, we should neglect both terms Eq.~(\ref{App32}) and
Eq.~(\ref{CPF3}); they are of the second order in $\Delta$. So,
Eq.~(\ref{Entro_Markov2}) is proven for $L=N-1$. By induction, the
equation can be written for arbitrary $L<N$.

\end{document}